\documentclass[10pt]{iopart}
\usepackage{graphicx}
\usepackage{epstopdf}
\usepackage{tikz}
\begin{document}
	
	%\pdfminorversion=4 
	
	\title[GGR method for DOS calculation]{Generalized Gilat-Raubenheimer method for density-of-states calculation in photonic crystals}
	
	%\author{Boyuan Liu,  Steven G. Johnson,  John D. Joannopoulos, Ling Lu*}
	%\address{}
	%\ead{}
	%\vspace{10pt}
	\author{Boyuan Liu$^1$, Steven G. Johnson $^2$, John D. Joannopoulos $^3$, Ling Lu$^1$}
	
	\address{$^1$Institute
		of Physics, Chinese Academy of Sciences/Beijing National Laboratory for Condensed Matter Physics, Beijing 100190, China}
	
	\address{$^2$Department of Mathematics, Massachusetts Institute of Technology, Cambridge, Massachusetts 02139, USA}
	
	\address{$^3$Department of Physics, Massachusetts Institute of Technology, Cambridge, Massachusetts 02139, USA}
	
	\ead{linglu@iphy.ac.cn}
	
	\begin{abstract}
		
		Efficient numeric algorithm is the key for accurate evaluation of density of states~(DOS) in band theory.
		Gilat-Raubenheimer~(GR) method proposed in 1966 is an efficient linear extrapolation method which was limited in specific lattices.
		Here, using an affine transformation, we provide a new generalization of the original GR method to any Bravais lattices and show that it is superior to the tetrahedron method and the adaptive Gaussian broadening method.
		Finally, we apply our generalized GR~(GGR) method to compute DOS of various gyroid photonic crystals of topological degeneracies.
	\end{abstract}
	
	\noindent{\it Keywords\/}: density of states, photonic crystal, topological photonics
	%\maketitle
	
	\ioptwocol
	
	\section{Introduction}\label{sec:intro}
	Numerical methods of DOS calculations~\cite{morris2014optados} fall into two categories, extrapolation and interpolation. Each category can use linear or high-order fittings.
	Linear extrapolation methods include GR~\cite{Gilat1966Accurate,Raubenheimer1967Accurate,kam1968accurate,finkman1971accurate,bross1993efficiency} and adaptive (Gaussian) broadening~\cite{yates2007spectral}.
	The high-order extrapolation methods were discussed in \cite{pickard1999extrapolative,pickard2000second}.
	Linear interpolation methods include the tetrahedron method~\cite{lehmann1972numerical,jepson1971electronic,blochl1994improved}, which does not need group-velocity information and is flexible in terms of volume grid division into tetrahedrons. The high-order interpolation methods were discussed in reference \cite{methfessel1983analytic,boon1986singular,methfessel1987singular}.
	
The extrapolation methods are better than the interpolation methods at band crossings~\cite{pickard1999extrapolative,muller1984band}.
The interpolation methods interpolate the frequency~(or energy) data from the nearest-neighbor momenta for linear interpolations and requires more neighboring data points for high-order interpolations. At the band crossings, interpolation methods sample the points across the degeneracy, resulting in the increase of errors. In contrast, an extrapolation method extrapolates the neighboring frequency data using both the frequency and the group velocity~(first derivative) at each momentum point for linear extrapolations and requires higher-order derivatives for high-order extrapolations. 
Consequently, the extrapolation methods are not vulnerable to the band crossings while the interpolation methods are.
	
GR is the first linear extrapolation method proposed. It was originally formulated in the three-dimensional~(3D) cubic grid and was extended to hcp~\cite{Raubenheimer1967Accurate}, tetragonal~\cite{kam1968accurate} and trigonal lattices~\cite{finkman1971accurate}, by dividing the irreducible Brillouin zones~(IBZ) into rectangular and triangular prisms.
An improved GR method~\cite{bross1993efficiency} derives the analytical formulation of the DOS contribution for parallelepiped subcells, applicable to all Bravais lattices.
In this work, using a geometric transformation between a cube and a parallelepiped, we made a simpler generalization of the original GR method for all lattices. The convergence plots show that our GGR method is consistently more accurate than the commonly-used tetrahedron and Gaussian methods.
In \ref{Append1}, we showed that this GGR method is equivalent to the improved GR method derived in a different way.
In \ref{Append2}, we discussed the GGR method for 2D.

%GR is the first linear extrapolation method proposed. It was originally formulated in cubic grid and was extended to hcp, tetragonal and trigonal lattices~\cite{Raubenheimer1967Accurate,kam1968accurate,finkman1971accurate}, by dividing the irreducible Brillouin zones~(IBZ) into rectangular and triangular prisms.An improved GR method~\cite{bross1993efficiency} derives the analytical formulation for the DOS contribution of parallelepiped subcell through the integration of delta functions, applicable to all Bravais lattices. In this work, we used a geometric transformation to generalize the original GR method to get an equivalent result as the improved GR method through the integration of group-velocity reciprocal on the equi-frequency surfaces. We use a simple affine transformation on wave vectors to transform the parallelepiped subcells in BZ into cubes, so that the original GR formulation can be used directly. We also compare our generealized GR method with two commonly used methods: tetrahedron method and adaptive Gaussian broadening method. The convergence plot shows that our method is the most accurate one.
	
In photonics, GR method has never been adopted. Tetrahedron method~\cite{busch1998photonic} and histogramming~\cite{johnson2002ultrafast,nikolaev2009accurate,kano2008analysis} were used instead. A new method named Dirichlet-to-Neumann maps~\cite{liu2011efficient} has been implemented in 2D photonic crystals for finding both the DOS and the equifrequency surfaces. In this paper, we applied the GGR method to photonic crystals.
	
	The rest of the paper is arranged in the following way.
Section~\ref{sec:GR} introduced the details of this transformation for our GGR method.
Section~\ref{sec:comp} compared the convergence of different methods.
In section~\ref{sec:gy}, we applied our method to topological photonic crystals.
Section~\ref{sec:time} discussed the computing efficiency of the GGR method.
Section~\ref{sec:conclusion} concluded our findings.
	
	\section{Generalizing GR method by affine transformation}\label{sec:GR}
	\begin{figure*}[ht]
		\includegraphics[width=\textwidth]{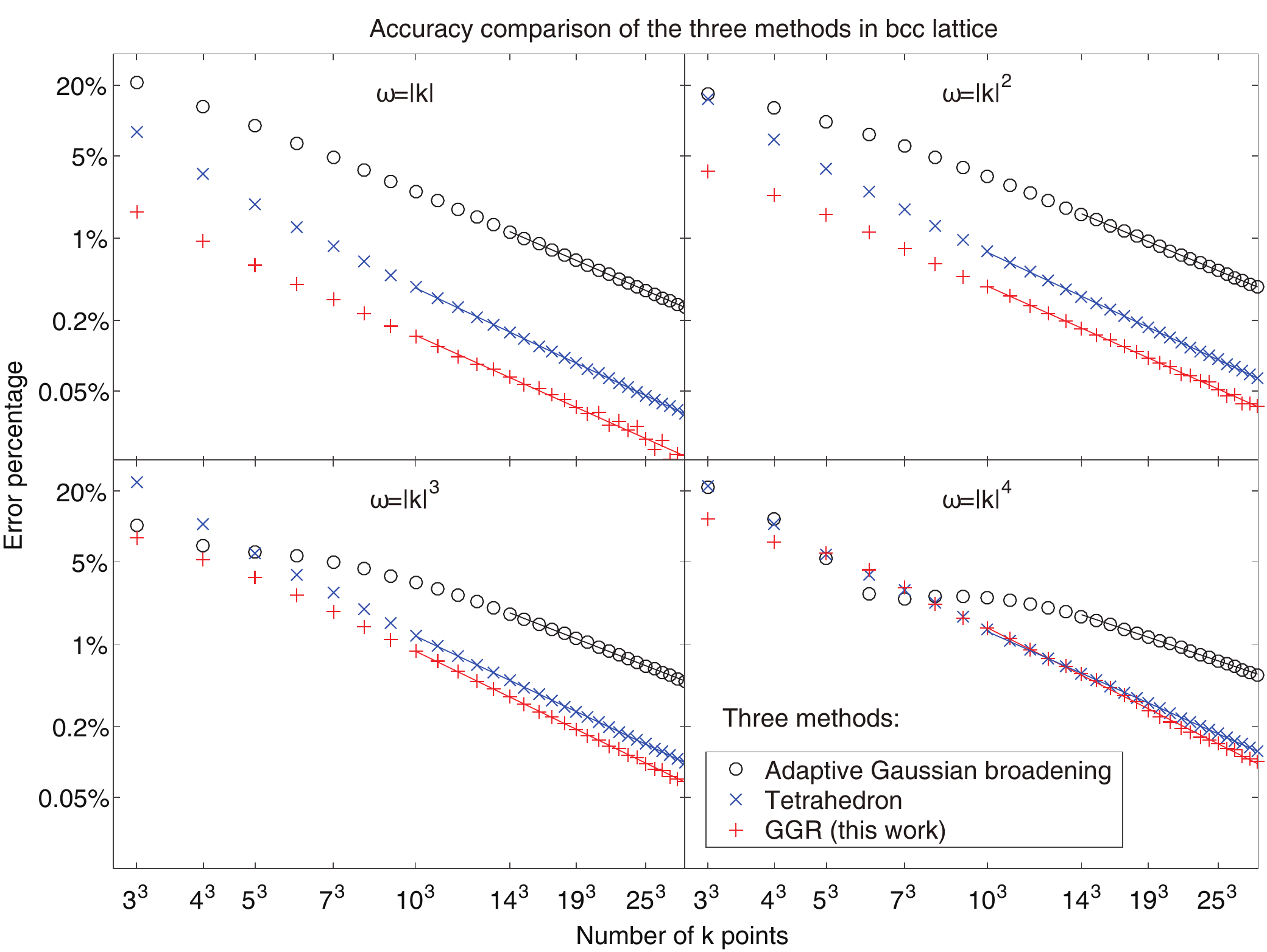}
		\caption{\label{fig:pc} \textbf{$\mathbf{error(N)}$ of the three methods in bcc lattice in double-logarithmic plots}. We assume that the band dispersion are $\omega = |\mathbf{k}|,   |\mathbf{k}|^2,  |\mathbf{k}|^3,  |\mathbf{k}|^4$ respectively. The data we adopt to line fitting is from $N=10^3$ to $32^3$ for GGR and tetrahedron method and is from $N=15^3$ to $32^3$ for adaptive Gaussian broadening method.
We note that the accuracy of tetrahedron method will be even worse in actual band structures with band crossings.}
	\end{figure*}
	
	The core idea of our GGR method is to use an affine transformation~\cite{bross1993efficiency} to transform a parallelepiped BZ into a cube, so that the original GR method can be used for any lattice. The BZ is a parallelepiped constituted by three reciprocal vectors $\mathbf{b}_i$ $(i = 1,2,3)$, starting from an arbitrary point $\mathbf{k}_0$.
The $\mathbf{k}$ points are uniformly distributed along three basis vectors $\mathbf{b}_i$.
The affine transformation changes the $\mathbf{k}$-basis of the parallelepiped BZ into $\mathbf{t}=(t_1, t_2, t_3)$ of a cubic volume, 
	
	\begin{equation}\label{eq:affine}
	\mathbf{k} - \mathbf{k}_0=B \mathbf{t}=\mathbf{b}_1 t_1+\mathbf{b}_2 t_2+\mathbf{b}_3 t_3,
	\end{equation}
	where  $ t_1,t_2,t_3 \in [0, 1]$ and $B = [\mathbf{b_1}, \mathbf{b_2}, \mathbf{b_3}]$. Consequently the volume elements of the two sets of bases satisfies $\rmd V_k=\det(B) \rmd V_t=\Omega \rmd V_t$, in which $\Omega$ is the volume of the BZ.
	
	We convert the DOS~[$D(\omega)$], the integral on equifrequency surface $S_\omega$, from the $\mathbf{k}$ basis into the cubic $\mathbf{t}$ basis
	\begin{equation}\label{eq:g}
	D(\omega) =\frac{1}{\Omega} \sum_n \int_{S_\omega} \frac{\rmd S_k}{|\mathbf{v_k}|}=\sum_n \int_{S_\omega} \frac{\rmd S_t}{|\mathbf{v_t}|},
	\end{equation}
	since	
	\begin{equation}\label{eq:g1}
	\frac{1}{\Omega}\frac{\rmd S_k\rmd k_\bot}{|\mathbf{v_k}|\rmd k_\bot}=\frac{1}{\Omega}\frac{\rmd V_k}{\rmd \omega}=\frac{\rmd V_t}{\rmd \omega}=\frac{\rmd S_t\rmd t_\bot}{|\mathbf{v_t}|\rmd t_\bot},
	\end{equation}
where $\mathbf{v_k}$ and $\mathbf{v_t}$ are the group velocities in each basis and $k_\bot$ and  $t_\bot$ are the vectors normal to $S_\omega$. $n$ is the band index.
$\mathbf{v_t}$ is obtained by scaling $\mathbf{v_k}$:
	\begin{equation}
	\mathbf{v_t}=\nabla_\mathbf{t} \omega(\mathbf{k(\mathbf{t})}) = \nabla_\mathbf{k} \omega \cdot \nabla_\mathbf{t} (B\mathbf{t})=\mathbf{v_k} \cdot B,
	\end{equation}
	where $\mathbf{v_k} \cdot B $ is a vector whose $i$th component is $(\mathbf{v_k} \cdot \mathbf{b}_i)$.  
	
	So far we have transformed the integral in parallelepiped BZ into integral in cubic volume $t_i \in [0,1]$. Then we can use original GR method to calculate the DOS in the basis of $t_i$. The GR method  partitions the cubic integral volume into uniform small cubes, with the $\mathbf{k}$ points at their centers. In each cubic subcell, we use linear extrapolation based on the frequency and group velocity of the central point to approximate the frequency of other region. In this case, the equifrequency surface of a given frequency is a polygon in each cubic cell. The area of the polygon is provided by the original GR method~\cite{Gilat1966Accurate}.
The final GGR formula is given in equation \ref{eq:10} in the Appendix.

%Take every subcell's velocity and the area of the polygon into definition (\ref{eq:g}), we can get the DOS contribution of all the subcells, which is equal to DOS of the initial untransformed photonic crystal system as shown in equation (\ref{eq:g2}).
	
	The integral region of our method is the whole BZ, a parallelepiped spanned by $\mathbf{b_1,b_2,b_3}$ or a fraction of it. This works for the general case for all lattices and symmetries. Although using IBZ is computationally more efficient, the choice of IBZ is symmetry specific. For example, we broke both the spatial and time-reversal symmetries in section~\ref{sec:gy}, and the corresponding IBZ varies from case to case. When sampling the whole BZ, we recommend that the $\mathbf{k}$ mesh be shifted away from the BZ center~($\Gamma$) to avoid the divergence problem due to the zero group velocities~\cite{Gilat1966Accurate}.

	\section{Accuracy comparison between three methods}\label{sec:comp}
	
	We compared the accuracy of GGR method with that of adaptive Gaussian broadening method and tetrahedron method. 
	We assume that the lattice is body-centered cubic~(bcc) and the BZ is a parallelepiped formed by three reciprocal lattice vectors $\mathbf{b}_i$ ($i = 1,2,3$ and $|\mathbf{b}_i|=1$) starting from origin. The total number of $\mathbf{k}$ points is $N = N_1 N_2 N_3$, where $N_i$ is the number of $\mathbf{k}$ points along $\mathbf{b}_i$ direction, and for simplicity, we set $N_1 = N_2 = N_3$. 
	The band frequency is $\omega = |\mathbf{k}|, |\mathbf{k}|^2, |\mathbf{k}|^3, |\mathbf{k}|^4$ respectively, so that we have analytical DOS to compare with.
The error percentage is defined as:
	\begin{equation}\label{eq:error}
	error(N)=\frac{\int_0^{1} |D_N(\omega)-D_\infty(\omega)| \rmd\omega}{\int_0^{1} D_\infty (\omega) \rmd\omega},
	\end{equation}
	where $D_N(\omega)$ is the DOS calculated on $N$ $\mathbf{k}$ points and $D_\infty(\omega)$ is the theoretical DOS.	
		
	In figure~\ref{fig:pc}, $error(N)$ of the three methods are presented in double logarithmic plots. The GGR method is better in the four cases.
It is important to point out that, in the realistic band structures with band crossings, the tetrahedron interpolation method will have an even lower accuracy~\cite{pickard1999extrapolative,muller1984band}. Therefore the GGR extrapolation method is a clear winner.
	
%The Gaussian method is known to be a primitive approach of low accuracy.
%The tetrahedron method is less accurate than the GGR method due to the lack of group-velocity information.
	We fit the errors linearly~[$ln(error(N)) = p_1 ln(N) + p_2$] for large number of $\mathbf{k}$ points,
	where $p_1$ and $p_2$ are the real fitting parameters.
	%, which give a power-law dependence: $error(N) = ln(p_2) N^{p1}$.
	The power dependences of $p_1$ were tabulated in table~\ref{tab:error} for all three methods.
	The $p_1$ values of GR method are consistent with the accuracy analysis in \cite{gilat1972analysis} which showed $error(N) \propto N^{-2/3}$. The $p_1$ values of the tetrahedron method are also close to the rate of convergence in \cite{wiesenekker1991quadratic}.
	
	\begin{table}[ht]
		\caption{\label{tab:error} The fitting parameters of adaptive Gaussian broadening,  tetrahedron and our GGR method in figure~\ref{fig:pc}.}
		\begin{tabular}{c|cccc}
			\hline
			$p_1$&　$\omega=|\mathbf{k}|$ & $\omega=\mathbf{k}^2$ & $\omega=|\mathbf{k}|^3$ &$\omega=\mathbf{k}^4$ \\
			\hline
			Gaussian&-0.6545&-0.6314&-0.5882&-0.5118\\
			\hline
			Tetrahedron&-0.7067 & -0.7059 & -0.7103 & -0.6712\\
			\hline
			GGR&-0.6786 & -0.6757 & -0.7353 & -0.7625 \\ 
			\hline
		\end{tabular}
	\end{table}
	
	% the time cost by group velocity hasn't been tested
	We wrote the GGR method according to reference~\cite{Gilat1966Accurate}, the adaptive Gaussian broadening method following reference~\cite{yates2007spectral}, and the tetrahedron method following reference~\cite{lehmann1972numerical,blochl1994improved}.
	In our program of the adaptive Gaussian broadening method, width of Gaussian function is $\alpha |\mathbf{v_k}| \Delta k$, where $\Delta k$ is the side length of a subcell. We set $\alpha=1.0$, which is a dimensionless constant indicating the broadening level.%$\alpha$ is a custom parameter and we set a reasonable value $\alpha = 1.0$~\cite{yates2007spectral}.
	We compared our GGR method program to the original GR method program ``GRINT" on CPC Program Library for simple cubic lattice.
	Our program of tetrahedron method was compared with the program ``tflovorn/ctetra" on github. In both methods, we got numerical consistence.

	\section{DOS of gyroid photonic crystals}\label{sec:gy}
	
\begin{figure*}[h]
 \includegraphics[width=\textwidth]{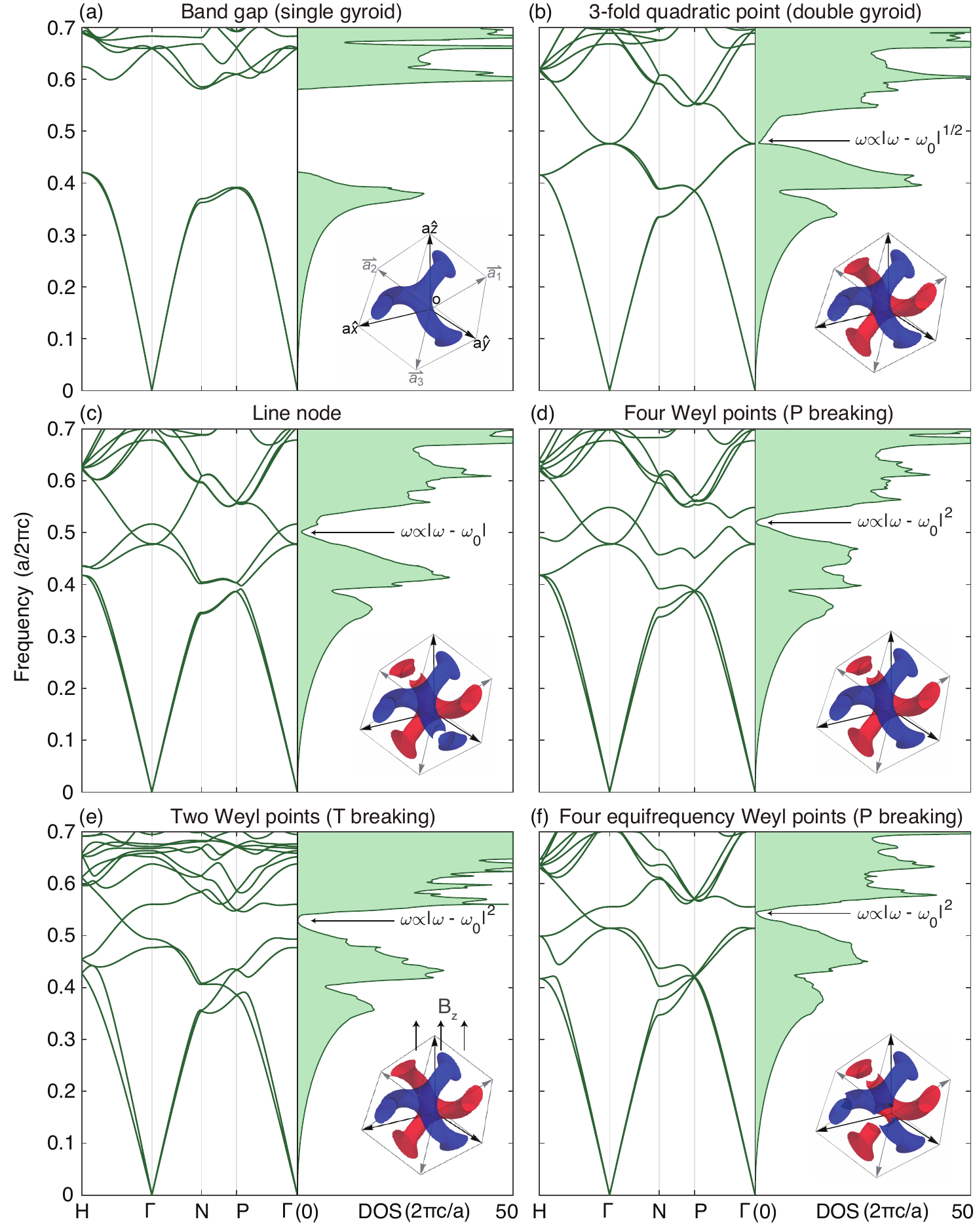}
 \caption{\label{fig:DOS}
 \textbf{DOS of six gyroid photonic crystals.}
Gyroid photonic crystal with a band gap (a), a quadratic degeneracy point (b), a line node (c) and Weyl points (d-f). 
The designs of (a-e) are from reference \cite{lu2013weyl} and the design of (f) is from reference \cite{Wang2015Topological}. Their dielectric constant is 16.  Each inset shows the unit-cell geometry of the crystal whose air-sphere defects are enlarged~($0.13a$) in the illustration for the easy of identification, where $a$ is cubic lattice constant. 
}
\end{figure*}

Using the GGR method, we computed the DOS of six gyroid photonic crystals in figure~\ref{fig:DOS}(a-f), following the original designs from reference~\cite{lu2013weyl} and reference~\cite{Wang2015Topological} in which the DOS data were not presented.
The insets are the real-space geometries in bcc unit cells.
The band structures were calculated using MPB~\cite{Johnson2001:mpb} for the frequencies and group velocities at $15^3$ uniformly-distributed $\mathbf{k}$ points in the whole BZ.
%The $\mathbf{k}$ grid avoids the $\Gamma$ point to circumvent the singularities of zero group velocity.

Figure~\ref{fig:DOS}~(a) is the single gyroid having a large band gap.
Figure~\ref{fig:DOS}~(b) is the double gyroid~(DG) having a threefold quadratic degeneracy. The DOS around the degeneracy point, of frequency $\omega_0$, shows a square-root relation of $D \propto |\omega - \omega_0|^{1/2}$. 
Figure~\ref{fig:DOS}~(c) is the perturbed DG having a nodal ring. The DOS around the degeneracy line shows approximately a linear relation of $D \propto |\omega - \omega_0|$.
Figure~\ref{fig:DOS}~(d) is the parity~($\mathcal{P}$)-breaking DG having two pairs of Weyl points.
Figure~\ref{fig:DOS}~(e) is the time-reversal~($\mathcal{T}$)-breaking DG having one pair of Weyl points.
Figure~\ref{fig:DOS}~(f) is the $\mathcal{P}$-breaking DG having  two pairs of Weyl points of the same frequency, in which the radius of the four defect air spheres is $r=0.09a$. 
The DOS around the above Weyl points all shows a roughly quadratic relation of $D \propto  |\omega - \omega_0|^2$.

\section{Computing efficiency} \label{sec:time}
Figure~\ref{fig:pc} shows that the extrapolative GGR method is more accurate than the interpolative tetrahedron method by utilizing the extra data of group velocities, which requires extra computing time.
Fortunately, the group velocities can be efficiently computed using the Hellman-Feynmann theorem $\frac{\partial \omega_\mathbf{k}}{\partial k_i} = \langle U_\mathbf{k} | \frac{\partial\hat{H}_\mathbf{k}}{\partial k_i}  | U_\mathbf{k} \rangle$, where $| U_\mathbf{k} \rangle$ is the periodic part of the Bloch wave function and  $\hat{H}_\mathbf{k}$ is the Hamiltonian operator of the system. Using MPB for example, the computation time for band dispersions with and without group velocities only differ by less than 2\%. 
We note that the total computing time is proportional to the number of $\mathbf{k}$ points $N$, in which
the time for computing DOS is negligible compared with the time for computing the band dispersions.

	\section{Conclusion}\label{sec:conclusion}
	
	In summary, we generalized GR method to all Bravais lattices using an affine transformation, which outperforms the tetrahedron and adaptive broadening methods. 
	Our GGR method divides BZ into parallelepipeds and such extrapolation method is advantageous in treating band crossings than interpolation methods.
	Future work includes high-order extrapolations~
	\cite{Saye2015High} and more versatile sub-cell division. 
	Our codes for the GGR and tetrahedron methods will be available for download at https://github.com/boyuanliuoptics/DOS-calculation.
	
	\ack
	
	We thank Tingtao Zhou for the initial efforts in this project and Hongming Weng and C.T. Chan for discussions. Boyuan Liu thanks Hao Lin and Qinghui Yan for their help on numerics. L.L. was supported by the National key R\&D Program of China under Grant No. 2017YFA0303800, 2016YFA0302400 and by NSFC under Project No. 11721404.
	J.D.J. and S.G.J. was partly supported
	by the Army Research Office through the Institute
	for Soldier Nanotechnologies under contract no.
	W911NF-13-D-0001.
	S.G.J. was supported in part by the Air Force Research Laboratory under Agreement No. FA8650-15-2-5220.
		\appendix

		\section{Equivalence between GGR and improved GR method} \label{Append1}
		%\subsection*{Equivalence between GGR and improved GR method}
		Here we prove that our GGR method is analytically equivalent to the improved GR method in reference~\cite{bross1993efficiency}.
		In the improved GR method, the DOS contribution of one subcell is given by 
		
		\begin{eqnarray}\label{eq:improGR}
		\fl
		\delta N =&- \frac{1}{2}\frac{1}{B_1B_2B_3}\sum_{\sigma_1=0}^1\sum_{\sigma_2=0}^1\sum_{\sigma_3=0}^1(-1)^{\sigma_1+\sigma_2+\sigma_3}\nonumber\\
		&(A-\sum_{i=1}^{3}(-i)^{\sigma_i}B_i)^2 \times \theta(A-\sum_{i=1}^{3}(-i)^{\sigma_i}B_i),
		\end{eqnarray}
		where $A=\omega-\omega_{\rm c} $, $B_i=\frac{1}{2} \mathbf{v} \cdot \mathbf{b_i} /N_i (i=1,2,3)$, $\theta(x)$ is the Heaviside step function, $\omega_{\rm c}$ is the frequency of central point of the subcell, $\mathbf{v}$ is the group velocity of this subcell, $\mathbf{b_i}$ is the reciprocal vector, $N_i$ is the number of $\mathbf{k}$ points along $i$th dimension and $N_1=N_2=N_3$.
		
		In order to compare the expression~(\ref{eq:improGR}) with that of our GGR method, we expand the above summation~(\ref{eq:improGR}):
		\begin{eqnarray}\label{eq:8}
		\fl	\delta N=-\frac{1}{2}\frac{1}{B_1B_2B_3} \nonumber\\
		\{[A-(B_1+B_2+B_3)]^2 \theta(A-(B_1+B_2+B_3))  \nonumber\\
		+ [A-(B_1-B_2-B_3)]^2 \theta(A-(B_1-B_2-B_3)) \nonumber\\
		+ [A-(-B_1+B_2-B_3)]^2 \theta(A-(-B_1+B_2-B_3)) \nonumber\\
		+ [A-(-B_1-B_2+B_3)]^2 \theta(A-(-B_1-B_2+B_3)) \nonumber\\
		- [A-(-B_1-B_2-B_3)]^2 \theta(A-(-B_1-B_2-B_3))\nonumber\\
		- [A-(-B_1+B_2+B_3)]^2 \theta(A-(-B_1+B_2+B_3)) \nonumber\\
		- [A-(B_1-B_2+B_3)]^2 \theta(A-(B_1-B_2+B_3)) \nonumber\\
		- [A-(B_1+B_2-B_3)]^2 \theta(A-(B_1+B_2-B_3))\}.
		\end{eqnarray}
		Without loss of generality, we assume $A>0$ and $B_1\geq B_2 \geq B_3\geq0$. Then the expression~(\ref{eq:8}) is transformed into a piecewise form,
		\begin{eqnarray}\label{eq:9}
		\fl
		\label{cases}
		\delta N=\cases{
			\frac{B_0}{B_1} \quad B_1\geq B_2 + B_3, 0\leq A\leq A_1\\
			\frac{1}{B_1B_2B_3}[2 (B_1B_2+B_2B_3+B_3B_1)\\
			\quad-(A^2+B_0^2)]\quad B_1\leq B_2 + B_3, 0\leq A\leq A_1\\ 
			\frac{1}{B_1B_2B_3}[ (B_1B_2+3B_2B_3+B_3B_1)\\
			\quad-A(-B_1+B_2+B_3)-\frac{1}{2}(A^2+B_0^2)]\\
			\qquad A_1\leq A\leq A_2\\ 
			\frac{2}{B_1B_2}[(B_1+B_2)-A]\quad A_2\leq A\leq A_3\\ 
			\frac{1}{2B_1B_2B_3}[ (B_1+B_2+B_3)-A]^2\\
			\qquad A_3\leq A\leq A_4\\
			0\quad A\geq A_4}
		\end{eqnarray}
		where $B_0 = (B_1^2+B_2^2+B_3^2)^{1/2}$, $A_1=|B_1-B_2-B_3|$, $A_2=(B_1-B_2+B_3)$, $A_3=(B_1+B_2-B_3)$, $A_4=(B_1+B_2+B_3)$.
		
		Next, we get the expression of DOS contribution of our GGR method according to section~\ref{sec:GR},
		\begin{eqnarray}\label{eq:10}
		\fl
		\label{cases}
		\frac{\rmd S_t}{|\mathbf{v_t}|} = \cases{
			\frac{4b^2}{v_{t1}}\quad v_{t1}\geq v_{t2} + v_{t3}, 0\leq \Delta\omega \leq \omega_1\\
			\frac{1}{v_{t1}v_{t2}v_{t3}}[2b^2(v_{t1}v_{t2}+v_{t2}v_{t3}+v_{t3}v_{t1})\\
			\quad-(\Delta\omega^2+(v_t b)^2)]\\
			\qquad v_{t1}\leq v_{t2} + v_{t3}, 0\leq \Delta\omega\leq \omega_1\\ 
			\frac{1}{v_{t1}v_{t2}v_{t3}}[b^2 (v_{t1}v_{t2}+3v_{t2}v_{t3}+v_{t3}v_{t1})\\
			\quad-b\Delta\omega (-v_{t1}+v_{t2}+v_{t3})\\
			\quad-\frac{1}{2}(\Delta\omega^2+(v_t b)^2)]\quad  \omega_1\leq \Delta\omega\leq \omega_2\\ 
			\frac{2}{v_{t1}v_{t2}}[b^2(v_{t1}+v_{t2})-v_t b\Delta\omega]\\
			\qquad \omega_2\leq \Delta\omega\leq \omega_3\\ 
			\frac{1}{2v_{t1}v_{t2}v_{t3}}[b (v_{t1}+v_{t2}+v_{t3})-\Delta\omega]^2\\
			\qquad \omega_3\leq \Delta\omega \leq \omega_4\\
			0\quad \Delta\omega\geq\omega_4
		}
		\end{eqnarray}
		where $\Delta \omega=\omega-\omega_{\rm c}$ and $v_t=|\mathbf{v_t}|$. $b=1/(2N_1)$ is half side length of subcell of the transformed cubic region. 
		Similarly, we assume that $\Delta \omega > 0$ and $v_{t1}\geq v_{t_2}\geq v_{t_3}\geq 0$, where $v_{ti} = \mathbf{v_k}\cdot \mathbf{b_i}$ is the component of transformed $\mathbf{v_t}$ ($i=1,2,3$). 
		And $\omega_1=b|v_{t1}-v_{t2}-v_{t3}|$, $\omega_2=b(v_{t1}-v_{t2}+v_{t3})$, $\omega_3=b(v_{t1}+v_{t2}-v_{t3})$, $\omega_4=b(v_{t1}+v_{t2}+v_{t3})$.
		
		The expressions of DOS calculation from one subcell (\ref{eq:9}) and (\ref{eq:10}) are equivalent. They only differ by a constant which is $\rmd S_t/|\mathbf{v_t}| =\delta N / (8N_1N_2N_3)$.
		
		%Now it is obvious that the $\delta N$~(\ref{eq:9}) is equivalent to $\rmd S_t/|\mathbf{v_t}|$~(\ref{eq:10}). The two expressions only differ by a constant, that is . 
	
	\section{GGR method in 2D} \label{Append2}
In order to use 3D GGR method for 2D lattices, we simply duplicate the frequency bands along a third imaginary dimension, so that the same GGR formulation applies with the following caveat.
%In 2D, $v_{t3} = 0$ for all the k points in the 3D BZ.

In 3D, the DOS formula~(\ref{eq:10}) is continuous (shown in figure~1 in \cite{Gilat1966Accurate}). However, for the extended 2D bands, the derivative of DOS is discontinuous due to $v_{t3} = 0$, $\omega_1=\omega_2$ and $\omega_3=\omega_4$. Thus, the 2D formula becomes,
	\begin{eqnarray}\label{eq:9}
	\fl
	\label{cases}
	\frac{\rmd S_t}{|\mathbf{v_t}|} = \cases{
			\frac{4b^2}{v_{t1}}\quad 0\leq \Delta\omega \leq \omega_1\\
			\frac{2}{v_{t1}v_{t2}}[b^2(v_{t1}+v_{t2})-v_t b\Delta\omega]
			\quad\omega_1\leq \Delta\omega\leq \omega_3 }
	\end{eqnarray}
	whose first derivative is discontinuous at $\Delta\omega=\omega_1$. This discontinuity and the vanishing quadratic terms~($\Delta\omega^2$) lead to a zigzag DOS plot.
The zigzag behavior also exists in tetrahedron method for the same reason, when being extended to 2D.
%Importantly, the zigzagged DOS does not indicate lower accuracies.

%Finally, we note that the We can obtain frequency bands of dense sampling $\mathbf{k}$ points in a short time and all the three methods for DOS calculation will reach satisfactory accuracy with the dense sampling $\mathbf{k}$ points.

	\section*{References}
	%\bibliography{DOS} 
	\bibliographystyle{unsrt}
	
\end{document}